\documentclass[pre,aps,preprint,showpacs]{revtex4}
\usepackage{revsymb,latexsym,amssymb,amsmath,comment}
\usepackage{graphicx,psfrag,subfigure,stmaryrd}

\newcommand{\exval}[1]{\mbox{$\langle \, {#1}\, \rangle$}}
\newcommand{\be}{\begin{equation}}
\newcommand{\ee}{\end{equation}}
\newcommand{\bel}[1]{\begin{equation}\label{#1}}
\newcommand{\bea}{\begin{eqnarray}}
\newcommand{\eea}{\end{eqnarray}}
\newcommand{\ba}{\begin{array}}
\newcommand{\ea}{\end{array}}
\newcommand{\bef}{\begin{figure}}
\newcommand{\ef}{\end{figure}}

\begin{document}

\author{Micheal Schulz}
\affiliation{University Ulm
D-89069 Ulm, Germany}
\email{michael,schulz@uni-ulm,de}
\author{Steffen Trimper}
\affiliation{Institute of Physics,
Martin-Luther-University, D-06099 Halle, Germany}
\email{steffen.trimper@physik.uni-halle.de}
\title{Modified Jarzynski Relation for non-Markovian noise }
\date{\today }
\begin{abstract}

We demonstrate the conventional Jarzynski relation (JR) is violated for a non-Markovian process with colored noise. 
As an example an exactly soluble model is considered with a simple protocol for the external work performed 
on the system along a non-equilibrium trajectory. For that model we derive an exact expression for the 
dissipative energy in terms of an  arbitrary correlator of the noise characterized by an  
autocorrelation time $t_c$. As the result we find corrections to the JR in terms of 
$t_c$. In the limiting case of a Gaussian process as well as an infinitely slow process the conventional JR is  
retained. The result is valid for an arbitrary colored noise.

\end{abstract}

\maketitle

\section{Introduction}

During the last decade, initiated by the seemingly remarkable relation derived in \cite{j}, a number of exact 
relations have been derived for non-equilibrium processes. The prominent equation in that context is the 
Jarzynski relation (JR) \cite{j}. This relation joins the change of the equilibrium free energy 
$\Delta F = F_f-F_i$ between two configurations of a system, an initial ($i$) and a final state ($f$), with 
the non-equilibrium work function $W$ spent in driving the system from the $i$ to the $f$ state:     
\be
\exval{e^{-\beta W}}  = e^{- \beta \Delta F}
\label{jr}
\ee
Here $\beta = (k_B T)^{-1}$ is the inverse temperature of the heat bath to that the system is coupled, 
$k_B$ is the Boltzmann constant. The relation is a highly non-trivial extension of the thermodynamic 
inequality $W \geq \Delta F$ valid for fixed temperature \cite{ll}. The substantial difference consists 
of the inclusion of an underlying dynamical process indicated by the average procedure in Eq.~\eqref{jr}. 
Originally the JR was derived using a deterministic dynamics based on a Hamiltonian formulation \cite{j}. 
Later the relation had been extended to a Langevin dynamics \cite{j1} or a discrete Markov equation 
\cite{c,c1}. The validity of the JR can be demonstrated for an exactly soluble models \cite{mj}. Recently the 
JR is also applied in single-molecule pulling experiments \cite{hs,ldstb,crjstb} in order to measure the 
folding free energies, and in calculations within the kinetic theory of gases \cite{lg,bbk}. Likewise, 
the non-equilibrium dynamics of gene expression and its relation to the JR had been studied in \cite{b}. 
An extension of the proof of the JR, solely based on specific equations of motion without any further assumption 
and non-Hamiltonian dynamics, has been proposed in \cite{cu}. Further, the relation between supersymmetry and 
non-equilibrium work relations was discussed in \cite{mmo}. Different aspect of the JR had been 
explored in several other papers such as in \cite{s}, where the entropy production along a stochastic trajectory had been 
analyzed and a generalized fluctuation theorem could be derived. Time symmetric fluctuations and their relation 
to the JR are obtained in \cite{mw}. JR in quantum systems had been studied in \cite{tlh,en}. A special problem 
discussed in that context is to find out an expression for the characteristic function of work performed on a 
quantum system \cite{thm}. Work fluctuations in quantum spin chains were studied recently \cite{dpk}. 

The broad spectrum of papers devoted to the JR and related problems suggest that the JR is generally valid. In the present 
paper we show for a simple example that the conventional JR is violated for a non-Markovian process. Especially, 
we calculate explicitly corrections to the JR in terms of a parameter which characterizes the non-Markovian 
behavior. Obviously one is confronted with different interpretations concerning the fact what a non-Markovian process means. 
Whereas a history dependent process with a white noise is classified as a Markovian process \cite{blp}, sometime a process with memory
and white noise is declared as a non-Markovian process \cite{ss}. In this paper we follow the classification 
scheme given by van Kampen \cite{vk}. In that context a process subjected to a colored noise is denoted as a non-Markovian process.  
For such kind of processes we demonstrate the breakdown of the JR. As shown directly the reason for the violation of conventional JR 
consists of the coupling of different time scales. In our approach there appears a term, by which the work protocol and the noise term 
are coupled. Whereas in case of a white noise the coupling between noise and working function is 
instantaneously, in case of a colored noise, the protocol is coupled to all former time scales 
of the noise. Eventually this coupling leads to a modification of the JR.

\section{Model}

To demonstrate the violation of the Jarzynski relation we consider the simple model of a damped oscillator which 
follows the equation for the variable $x(t)$
\be
m \ddot x + \gamma \dot x + m \omega_0^2 x = m [ K(t) + \eta(t) ] \,.
\label{evo}
\ee
As usual $m$ is the mass, $\gamma$ is a damping parameter and $\omega_0$ is the frequency of the undamped oscillator. 
The function $K(t) $ is an external force. For simplicity we assume the following protocol for that external force
\be
K(t; \tau) = \left \{\begin{array}{ccc} 
0 &  & -\infty < t \leq 0 \\
\displaystyle{\frac{K_0}{\tau}\, t} &  & 0 < t \leq \tau \\
K_0 &  & \tau < t < \infty \end{array} \right . 
\label{pro}
\ee
The last relation means that the system is initially in equilibrium. At time $t=0$ the system is perturbed by a 
time dependent force. After a time interval $\tau > 0$ the external force is switched of and the system tends 
again to an equilibrium state. Notice that the limiting case $\tau \to \infty$ characterizes an infinite slow process. 
During such a quasistatic process the systems remains in equilibrium. Aside from an external force the system 
should be subjected to a stochastic force $\eta(t)$ with the properties
\be
\langle\, \eta (t)\,\rangle = 0,\quad \langle \,\eta(t)\, \eta(t')\, \rangle = S(\,t - t'\,)\,.
\label{noi}
\ee
The probability measure corresponding to Eq.~\eqref{noi} is given by 
\be
P(\eta) = \mathcal{N} \exp\left ( - \int dt \int dt'\, \eta (t) S(t-t') \eta (t')\right)\,, 
\label{prob}
\ee
where $\mathcal{N}$ is a normalization factor. In case the correlator fulfills $S(t) \propto \delta (t)$,  
the process is a Gaussian process with a white noise. Rather, $\eta(t)$ is supposed to be 
a stationary process and the autocorrelation function $S(t)$ reflects a colored noise behavior. That fact offers 
some profound differences with the conventional Langevin equation with a white noise \cite{vk}. As for instance 
the autocorrelation time $t_c$ of $\eta(t)$ is not zero, the stochastic process governed the evolution equation for 
$x(t)$ is a non-Markovian one. Due to the correlator $S(t)$ the values of $\eta(t_1)$ for $t_1 > t_0$ 
with $\eta(t_0)$ is fixed, depends on all the values of $\eta(t)$ inside the interval $t_0 < t < t_1$. Although 
our result is independent on the special realization of $S(t)$, we consider to be specific the following colored noise behavior 
\be
S(t) = \frac{\gamma}{k_B T} \exp (\,-  \mid t \mid /t_c )\,.
\label{noi1}
\ee   
The system is assumed to be coupled to a heat bath with the fixed temperature $T$.
  
For the subsequent discussion we need the Greens function related to Eq.~\eqref{evo}:
\bea 
G(t-t') &=& \Theta (t - t') g( t - t'),\nonumber\\\mbox{with} \quad g(t-t') &=& \frac{e^{-\kappa (t - t')}}{\omega_e} \sin \omega_e (t - t'), \quad
\kappa = \frac{\gamma}{2m},\quad \omega_e = \sqrt{\omega_0 ^2 - \kappa^2} \,. 
\label{gf}
\eea
In terms of the Greens function the solution of Eq.~\eqref{evo} reads
\be
x(t) = \int_{-\infty}^{\infty} dt' G(t -t') [ \eta (t') + K(t') ] \,.
\label{sol}
\ee
The work performed on the system up to time $t$ is given by 
\be
W(t;\tau, t_c) = \int_{-\infty} ^t dt' K(t') \dot x(t') \,. 
\label{sol1}
\ee
Apparently the work depends on both the time $\tau$, compare Eq.~\eqref{pro}, and on the correlation time $t_c$ 
of the colored noise according to Eq.~\eqref{noi1}. 

Due to the initial condition $G(t=0) = 0$ it results
\be
\dot x(t) = \int_{-\infty}^t dt' \frac{d}{dt} g(\,t -t'\,) [\, \eta (t') + K(t')\, ] \,.
\label{sol2}
\ee
We are interested in the work performed during the entire interval $-\infty < t < \infty$.  
Inserting Eq.\eqref{sol2} into Eq.~\eqref{sol1} the work function can be expressed by 
$ W(\infty;\tau,t_c) \equiv W_\eta (\tau;t_c) + W_K (\tau)$ with  
\bea   
W_\eta (\tau,t_c) &=& \int_{-\infty}^{\infty} dt' K(t',\tau) \int _{\infty}^{t'} dt''\,\frac{d}{dt'} g (\, t' - t''\,) \eta (t'')\nonumber\\    
W_K (\tau) &=& \int_{-\infty}^{\infty} dt' K(t',\tau) \int _{\infty}^{t'} dt''\,\frac{d}{dt'} g (\, t' - t''\,) K (t'') \,.  
\label{sol3}
\eea
Whereas the work due to the external force, denoted as $W_K(\tau)$, depends only on the duration time $\tau$, see Eq.~\eqref{pro}, 
the stochastic part $W_\eta(\tau,t_c)$ is influenced by 
both time scales $\tau$ and $t_c$.  In particular, the first relation of Eq.~\eqref{sol3} offers a multiplicative coupling between the 
work protocol $K(t;\tau)$ and the noise $\eta(t)$. While in case of a white noise both quantities $K(t)$ and $\eta(t)$  
are instantaneously coupled at the time $t$, in the case of a colored noise the work is coupled to 
the whole history of the noise. Eventually this coupling leads to the breakdown of the JR for the system driven by the colored 
noise. In performing the integration in Eq.~\eqref{sol3} we have to take into account the 
different time behavior of the external force according to Eq.~\eqref{pro}. It results
\bea 
W_\eta (\tau,t_c) &=& \int_{-\infty}^{\infty} dt R(t;\tau) \eta(t) \quad\mbox{with}\nonumber\\
R(t, \tau) &=& - \frac{K_0}{\tau} \Theta ( \tau - t) \int_{\mbox{max} (0, t)}^\tau dt' G(t'- t)\,.
\label{sol4}
\eea
For the subsequent calculation the Fourier transformed function $R(\omega)$ is required which is given by 
\be
R(\omega) = \frac {K_0 (e^{i\omega \tau}-1)}{2 \omega_e \omega \tau} 
\left[ \frac{\kappa - i( \omega - \omega_e)}{\kappa^2 + (\omega - \omega_e )^2} - 
\frac{\kappa - i( \omega + \omega_e)}{\kappa^2 + (\omega + \omega_e )^2}
\right]\,.
\label{sol5} 
\ee
Using the working protocol we can calculate the work function $W_K(\tau)$ according to Eq.~\eqref{sol3}. The result is
\bea
W_K (\tau ) &=& \frac{K_0^2 m}{\omega_0^4 \tau^2} \left\{1 - 4 \left(\frac{\kappa}{\omega_0}\right)^2 + 2 \kappa \tau + \frac{1}{2} (\omega_0 \tau )^2 
+ \right . \nonumber\\
&-&  \left . e^{-\kappa \tau} \cos(\omega_e \tau) \left[1 - 4 \left(\frac{\kappa}{\omega_0}\right)^2 \right] - \frac{ \kappa\,e^{-\kappa \tau } \sin (\omega_e \tau )}{\omega_e}
\left[ 3 - 4 \left(\frac{\kappa}{\omega_0}\right)^2 \right] \right\} \,.
\label{noi3a}
\eea
Now we are able to study in detail the behavior of the system.

\section{Generalized Jarzynski relation}

\begin{figure}[h]
\centering
\includegraphics[width=1.0\textwidth]{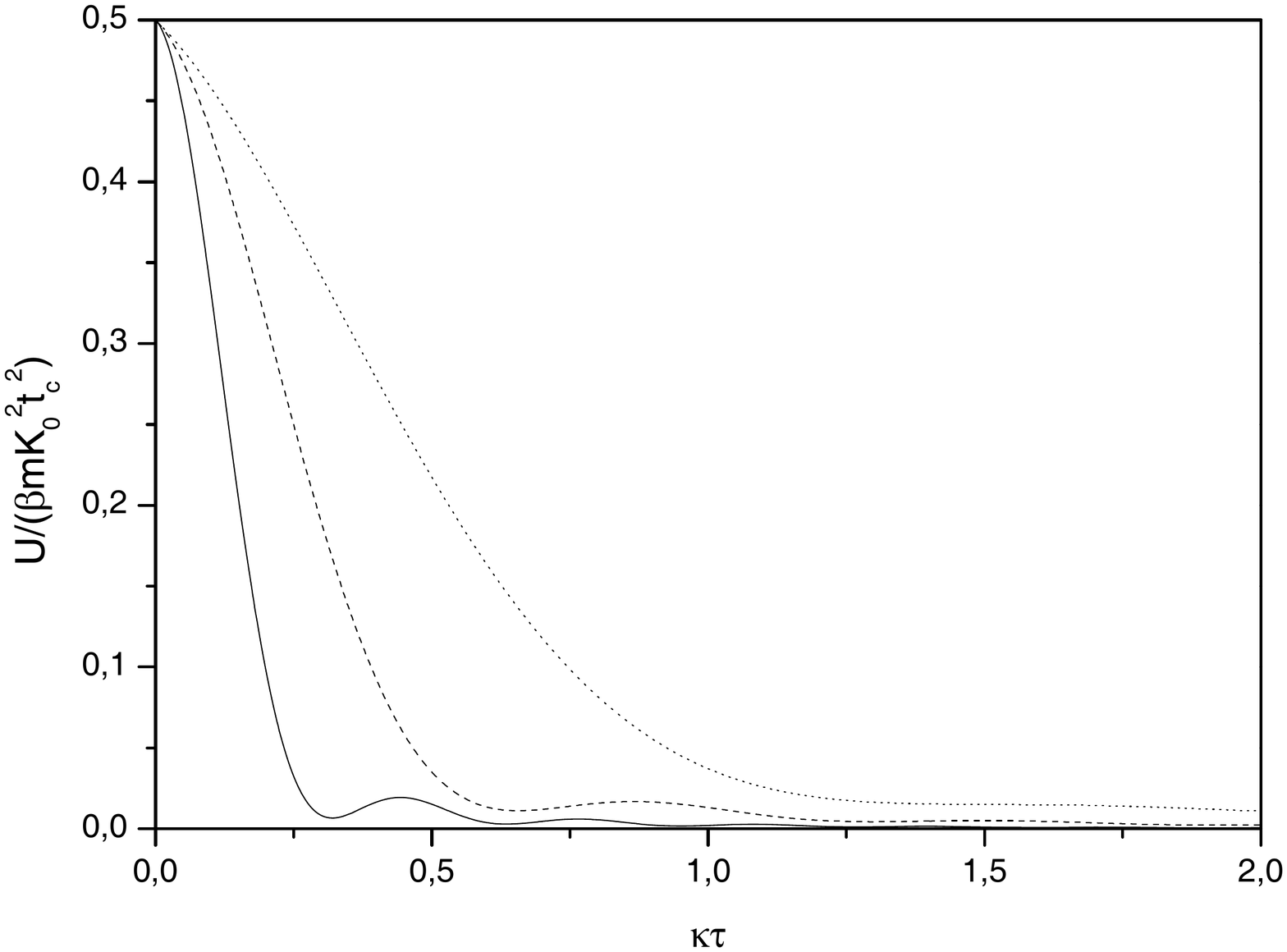}
\caption{Function $U$ according to Eq.~\eqref{vio} as function of $\kappa \tau$ for different ratios $\frac{\omega_e}{\kappa}$ = 
$20$ (solid); $10$ (dashed) and $5$ (dotted line)\,. }
\label{fig1}
\end{figure}

According to Eq.~\eqref{sol1} the total work is defined by 
\be
W(\infty, \tau ) = \int_{-\infty}^{\infty} dt R(t) \eta(t) + W_K(\tau ) \,. 
\label{w}
\ee
Here the function $R(t)$, see Eq.~\eqref{sol4}, can be rewritten as 
\be
R(t) = \int_t^{\infty} K(t') \partial_{t'} G(t'-t) dt' \,.
\label{w1}
\ee
Following Eq.~\eqref{jr} let us consider the expression
\be
\langle e^{-\beta W(\infty;\tau)} \rangle = \langle e^{- \beta \int_{-\infty}^{\infty} dt R(t) \eta(t)} 
\rangle\, e^{- \beta W_K(\tau )}\,. 
\label{w2}
\ee
Using Eq.~\eqref{prob} the average in the last equation can be easily performed. To that aim we remark that after   
making Fourier transformation the exponential term in the probability measure Eq.~\eqref{prob} reads
$$
\int_{-\infty}^{\infty} \frac{d\omega}{2 \pi} \eta* (\omega) S(\omega) \eta (\omega)\,.
$$
Performing the integration over the probability distribution we obtain
\be
\langle \exp(-\beta W(\infty;\tau,t_c)\rangle = 
\exp\left( \frac{\beta^2}{2 \pi} \int_{-\infty}^{\infty} d\omega\, \frac{\mid R(\omega)\mid\,^2\,}{S(\omega)} \right)\,
\exp(-\beta W_K(\tau)) \,.
\label{prob1}
\ee   
The correlation function for the colored noise, defined in Eq.~\eqref{noi1}, reads after Fourier transformation 
\be
S(\omega) = \frac{\beta \gamma}{ 1 + (\omega t_c)^2} \,.
\label{noi2}
\ee
In the limiting case $t_c = 0$ the Gaussian noise behavior is retained. 
Using the last expression combined with Eq.~\eqref{sol5} we get the contribution from the noise in the following form  
\bea
\int_{-\infty}^{\infty} \frac{d\omega}{2\pi}\, \frac{\mid R(\omega)\mid\,^2\,}{S(\omega)} &=& 
\frac{K_0^2 m }{\beta \omega_0^4 \tau^2} \left\{ 1 + 2 \kappa \tau + (\omega_0 t_c)^2 - 4 \left(\frac{\kappa}{\omega_0}\right)^2 
+ \right . \nonumber\\
&-& \left . e^{-\kappa \tau} \cos(\omega_e \tau) [ 1 + (\omega_0 t_c)^2 - 4 \left(\frac{\kappa}{\omega_0}\right)^2 ] \right . \nonumber\\
&-& \left . \frac{ \kappa\,e^{-\kappa \tau } \sin (\omega_e \tau )}{\omega_e}  
\left[  \,3 + (\omega_0 t_c)^2 - 4 \left(\frac{\kappa}{\omega_0}\right)^2 \right] \right\} \,.
\label{noi3}
\eea
To discuss the JR let us introduce the function $V(\tau,t_c)$ according to   
\be
\langle \exp(-\beta W(\infty;\tau,t_c)\rangle \equiv \exp(\beta V(\tau,t_c))\quad\rm{with}\quad 
V(\tau,t_c) = \beta \int_{\infty}^{\infty} \frac{d\omega}{2\pi} \frac{\mid R(\omega)\mid\,^2\,}{S(\omega)} -  W_K(\tau)\,.
\label{noi4}
\ee
In case the JR is fulfilled this quantity should be identified with the free energy $V(\tau,t_c) = -\Delta F$. 
Inserting Eqs.~\eqref{noi3} and \eqref{noi3a} into Eq.~\eqref{noi4} it results
\be
V(\tau,t_c) = \frac{ K_0^2 m }{2 \omega_0^2 } \left[ 2 \left(\frac{t_c}{\tau}\right)^2 
[\, 1 - e^{-\kappa \tau} \cos(\omega_e \tau) -  \frac{\kappa}{\omega_e} e^{-\kappa \tau} \sin(\omega_e \tau) \,]  - 1 \right] \,. 
\label{noi5}
\ee
As usually assumed the system is in equilibrium before the external force $K(t)$ is applied. After fixing the force $K(t) = K_0$ for 
$t \geq  \tau$ the system is again in an equilibrium state. Therefore we can calculate the difference of the free energy. 
The partition function is given by 
\be
Z(K) = \frac{1}{h^3} \int dx dp e^{-\beta H} \,.
\label{zu}
\ee 
From here we get the difference of the free energy between the equilibrium initial and final state
\be
\Delta F =  \beta^{-1} \ln \frac{Z(K_0)}{Z(0)} = \frac{K_0^2 m }{2\omega_0^2}
\label{fe}
\ee
Notice that the sign of $\Delta F$ is chosen in such a manner that it corresponds to the free energy of the damped oscillator 
applied to the system from outside.  To illustrate the behavior of our system let us introduce a new function $U$ by 
\be
U(\tau, t_c) = V(\tau,t_c) + \Delta F\,.
\label{vio}
\ee
If this function $U$ is nonzero for any finite $\tau$ and $t_c \neq 0$ the JR has to be modified.  
Inserting Eq.~\eqref{noi5} and Eq.~\eqref{fe} it results
\be
U(\tau,t_c) = \frac{ K_0^2 m }{\omega_0^2} \left(\frac{t_c}{\tau}\right)^2 
\left[ 1 - e^{-\kappa \tau} \cos(\omega_e \tau)  - \frac{\kappa}{\omega_e} e^{-\kappa \tau} \sin(\omega_e \tau) \right] \,.
\label{vio1}
\ee
The function $U(\tau,t_c) \geq 0$ is a measure for the dissipative energy. In case the correlation time $t_c$ of the 
colored noise tends to zero we find $U(\tau,t_c=0) = 0$. That case corresponds to a white noise behavior and 
consequently the JR is retained. Likewise for an infinitely slow process $\tau \to \infty$ it results 
$U(\tau \to \infty, t_c) = 0$ and the the JR is fulfilled again. In the last case the system remains in equilibrium. 
The behavior of the dissipation $U$ is shown in Fig.~\ref{fig1}.  
Notice that the ratio $\frac{t_c}{\tau}$ plays an decisive role for 
the breakdown of the JR. The JR is fulfilled for $\frac{t_c}{\tau} \to 0$, i.e. the characteristic memory time 
of the colored noise process should decay within the time of the work protocol. However in all other case, especially for 
a finite correlation time $t_c$, the conventional JR relation is violated. As argued before in that case the work performed on the 
system becomes history dependent and a more general relation is fulfilled:
\be
\exval{e^{-\beta ( W - \Delta F )}}  = e^{\beta U(\tau,t_c)} \,.
\label{jrn}
\ee
For a non-Markovian process the conventional JR is modified by the correlation time of the colored noise. 
Let us stress again that our result is not restricted to the special choice of the colored noise due to Eq.~\eqref{noi1} with 
a correlation time $t_c$. From Eqs.~\eqref{prob1} and \eqref{noi2} one concludes that for a Gaussian process $S(\omega)$ is 
independent on $\omega$ and consequently the JR is fulfilled. In case $S(\omega)$ depends on the frequency $\omega$ the JR 
has to be modified accordingly.

\section{Conclusions}
 
In this paper we have demonstrated that the conventional Jarzynski relation, which is broadly discussed in 
a variety of different fields, has to be corrected if the system is subjected to a colored noise. For such a 
realization the history dependent stochastic process offers a significant influence on the work performed on 
the system. Additionally to the fact that the work is not a state variable as already in the conventional 
statistical mechanics, the work function becomes even history dependent leading to additional contributions. 
Originated by the coupling of work function and noise the correction term to the JR is determined by the 
autocorrelation time of the colored noise. In view of our result the proof of the validity of the Jarzynski 
in \cite{ss} is restricted to a special class of systems with memory and not valid for all kinds of non-Markovian 
processes. Concerning that problem we are aware that there exits different interpretations. In the stronger classification 
proposed by van Kampen \cite{vk} systems with memory and white noise should be considered as Markovian. 
Here we have considered a simple model of non-interacting damped oscillators subjected to a colored noise which is  
doubtless a non-Markovian process.\\

This work has been supported by the DFG (SFB 418). M.S. is grateful to the University Halle 
for kind hospitality.  


\newpage
\begin{thebibliography}{90}

\bibitem{j} C.Jarzynski, Phys.Rev.Lett. {\ bf 78}, 2690 (1997)\,.

\bibitem{ll} L.D.Landau and E.M.Lifshitz, {\em Statistical Physics} (Pergamon Press,Oxford, 1990)\,.

\bibitem{j1} C.Jarzynski, Phys.Rev. E {\ bf 56}, 5018 (1997)\,.

\bibitem{c} G.E. Crooks, J.Stat.Phys. {\bf 90}, 1481 (1998); Phys.Rev. E {\ bf 60}, 2721 (1999)\,.

\bibitem{c1} G.E. Crooks, Phys.Rev. E {\ bf 61}, 2361 (2000)\,.

\bibitem{mj} O.Mazonka, C.Jarzynski, arXiv.cond-mat/9912121\,.

\bibitem{hs} G. Hummer, A. Szabo, Proc.Nat.Acad.Sci. USA {\bf 98}, 3658 (2001)\,.

\bibitem{ldstb} J.Liphardt, S.Dumont, S.B. Smith, I. Tinoco, and C. Bustamante, Science {\bf 296}, 1832 (2002)\,.

\bibitem{crjstb} D. Collin, F. Rotort, C.Jarzynski, S.B. Smith, I. Tinoco, and C. Bustamante, Nature {\bf 437}, 231 (2005)\,.

\bibitem{lg} R.C. Lua, A.Y.Grosberg, J.Phys.Chem. B {\bf 109}, 6805 (2005)\,.

\bibitem{bbk} I. Bena, C. Van den Broeck, and R.Kawai, Euro.Phys.Lett. {\ bf 71}, 879 (2005)\,.

\bibitem{b} J. Berg, Phys. Rev.Lett. {\bf 100}, 188101 (2008)\,.

\bibitem{cu} M. A. Cuendet, Phys.Rev.Lett. {\bf 96}, 120602 (2006)\,.

\bibitem{mmo} K. Mallick, M. Moshe, and H. Orland, ArXiv.cond-mat.: 0711.2059v1 (2007)\,.

\bibitem{s} U.Seifert, Phys. Rev.Lett. {\bf 95}, 040602 (2005)\,.

\bibitem{mw} C.Maes and M.H.van Wieren, Phys. Rev.Lett. {\bf 96}, 240601 (2006)\,.

\bibitem{tlh} P.Talkner, E.Lutz, and P.H\"anggi, Phys.Rev. E {\bf 75} 050102(R) (2007)\,

\bibitem{en} A.Engel and R.Nolte, Euro.Phys.Lett. {\bf 79}, 10003), (2007)\,.

\bibitem{thm} P.Talkner, P.H\"anggi, and M.Morillo, Phys.Rev. E {\bf 77}, 051131 (2008)\,.

\bibitem{dpk} S.Dorosz, T.Platini, and D.Karevski, Phys.Rev. E {\bf 77}, 051120 (2008)\,.

\bibitem{blp} G.Bussi, A.Laio, and M.Parrinello, Phys. Rev.Lett. {\bf 96}, 090601 (2006)\,.

\bibitem{ss} T. Speck and U. Seifert, J.Stat.Mech. L09002 (2007)\,.

\bibitem{vk} N. G. van Kampen, {\em Stochastic Processes in Physics and Chemistry} (North-Holland, 
Amsterdam, 1992). 

\newpage




\end{thebibliography}
\end{document}